\documentclass[12pt]{article}

\usepackage{textgreek}
\usepackage{amsmath,amssymb,graphicx}
\usepackage{color}
\usepackage{lineno}

\definecolor{brilliantrose}{rgb}{1.0, 0.33, 0.64}
\definecolor{byzantine}{rgb}{0.74, 0.2, 0.64}

\newcounter{lastnote}

\title{Experimental Evidence of Quantum Radiation Reaction in Aligned Crystals}

\author
{Tobias N. Wistisen,$^{1\ast}$ Antonino Di Piazza,$^{2}$ Helge V. Knudsen,$^{1}$ Ulrik I. Uggerh\o j$^{1}$\\
\\
\normalsize{$^{1}$Department of Physics and Astronomy, Aarhus University,}\\
\normalsize{Ny Munkegade 120, Aarhus, 8000, Denmark}\\
\normalsize{$^{2}$Max Planck Institute for Nuclear Physics,}\\
\normalsize{Saupfercheckweg 1, Heidelberg, 69117, Germany}\\
\\
\normalsize{$^\ast$To whom correspondence should be addressed; E-mail:  tobiasnw@phys.au.dk.}
}

\date{\today}

\begin{document} 


\baselineskip24pt


\maketitle


\newpage

\textbf{
Radiation reaction is the influence of the electromagnetic field 
emitted by a charged particle on the dynamics of the particle itself.
Taking into account radiation reaction is essential for the correct description of the motion of high-energy particles driven by strong electromagnetic fields. Classical theoretical approaches to radiation reaction lead to physical inconsistent equations of motion. A full understanding of the origin of radiation reaction and its consistent description are possible only within the more fundamental quantum electrodynamical theory.
However, radiation-reaction effects have never been measured, which has prevented a complete understanding of this problem.
Here we report experimental radiation emission spectra from ultrarelativistic positrons in silicon in a regime where both quantum and radiation-reaction effects dominate the dynamics of the positrons. We found that each positron emits multiple photons with energy comparable to its own energy, revealing the importance of quantum photon recoil. Moreover, the shape of the emission spectra indicates that photon emissions occur in a nonlinear regime where positrons absorb several quanta from the crystal field.
Our theoretical analysis shows that only a full quantum theory of radiation reaction is capable of explaining the experimental results, with radiation-reaction effects arising from the recoils undergone by the positrons during multiple photon emissions.
This experiment is the first fundamental test of quantum electrodynamics in a new regime where the dynamics of charged particles is determined not only by the external electromagnetic fields but also by the radiation-field generated by the charges themselves. Future experiments carried out in the same line will be able to, in principle, also shed light on the fundamental question about the structure of the electromagnetic field close to elementary charges.
}

A complete understanding of the dynamics of charged particles
in external electromagnetic fields is of fundamental importance
in several branches of physics, spanning e.g. from pure
theoretical areas like particle physics, to more applicative 
ones like accelerator physics. Since accelerated charges, 
electrons for definiteness, emit electromagnetic radiation, 
in the realm of classical electrodynamics a self-consistent 
equation of motion of the electron in an external electromagnetic 
field must take into account the resulting loss of energy and momentum 
\cite{Jackson_b_1975,Landau_b_2_1975}. 
However, the inclusion of the reaction of the radiation emitted 
by an electron on the motion of the electron itself 
(radiation reaction) leads to one of the most famous and 
controversial equations of classical electrodynamics, 
the Lorentz-Abraham-Dirac (LAD) equation \cite{Abraham_b_1905,Lorentz_b_1909,Dirac_1938,Barut_b_1980,Rohrlich_b_2007}. 
The LAD equation is plagued by serious inconsistencies
like the existence of ``runaway'' solutions, with the
electron acceleration diverging exponentially in time even if
the external field identically vanishes.
The mentioned exponential growth
is found to occur at time scales of the
order of the time that light needs to cover the classical
electron radius $r_e=e^2/mc^2=2.8\times 10^{-15}\;\mbox{m}$, i.e.
$\tau=r_e/c=9.5\times 10^{-24}\;\mbox{s}$ 
\cite{Jackson_b_1975,Landau_b_2_1975,Rohrlich_b_2007}. Now,
the classical electron radius is $\alpha=e^2/\hbar c\simeq1/137$ 
times smaller than the reduced Compton wavelength 
$\lambda_C=\hbar/mc=3.9\times 10^{-13}\;\mbox{m}$, 
which is the typical length of quantum electrodynamics 
(QED) \cite{Landau_b_4_1982} and this also applies to the time scale
$\tau$. This occurrence suggests that a complete
understanding of the problem of radiation reaction requires
an approach based on QED. Below we will concentrate on
a regime where both quantum and radiation-reaction effects
are substantial, and we therefore refer the reader to the recent 
reviews \cite{Hammond_2010_b,Di_Piazza_2012,Burton_2014} concerning
classical aspects of radiation reaction. We only recall here 
that the discussed overlapping between classical and quantum 
scales implies that within the realm of classical electrodynamics 
the LAD equation can be consistently approximated by an equation, 
known as Landau-Lifshitz (LL) equation, which is not plagued by
the above-mentioned inconsistencies \cite{Landau_b_2_1975}. 
In order for quantum effects to be negligible, in the instantaneous 
rest frame of a relativistic electron the external field has to vary
slowly on space (time) distances of the order of $\lambda_C$
($\lambda_C/c$) and its amplitude has to be much smaller than the so-called
critical electric (magnetic) field of QED $E_{cr}=m^2c^3/\hbar|e|=1.3\times 10^{18}\;\mbox{V/m}$ ($B_{cr}=m^2c^3/\hbar|e|=4.4\times 10^{9}\;\mbox{T}$)
\cite{Landau_b_4_1982,Dittrich_b_1985,Fradkin_b_1991,Baier_b_1998}.
These conditions guarantee, among others, that the recoil undergone by the electron 
during the emission of a single photon inside the considered field, 
is much smaller than the electron energy and the whole emission 
process can be treated classically as a continuous emission. 
In the situation investigated below, the second mentioned 
condition plays a major role and, if we define the parameter 
$\chi=\gamma E/E_{cr}$, where $E$ is a measure of the amplitude 
of the crystal electric field and where $\gamma$ is the electron 
Lorentz factor, it can be expressed as $\chi\ll 1$. The 
relation between radiation reaction and the emission of 
multiple photons in a regime where quantum recoil is substantial 
has been pointed out in \cite{PhysRevLett.105.220403} in the 
interaction of ultrarelativistic electrons with intense 
laser fields. In the quantum radiation reaction regime the typical 
energy of the emitted photons is comparable to the energy of the 
incoming electron ($\chi\gtrsim 1$) and multiple photon emission 
is more probable than single photon emission. In laser-electron interaction
it is customary to assume the laser field to be ``strong'' in the sense that the 
parameter $\xi_l=|e|E_l/mc\omega_l$, where $E_l$ and $\omega_l$ are 
the laser amplitude and the angular frequency, respectively, is much 
larger than unity \cite{PhysRevLett.112.015001,PhysRevLett.105.220403,PhysRevLett.111.054802}. At $\xi_l\gg 1$ each photon emission occurs with the absorption of several laser photons and the radiation is formed over a length (formation length) much shorter than the laser wavelength. Thus, the formula of the radiation emission probability in a constant crossed field (CCF) can be employed \cite{Baier_b_1998,Ritus,Landau_b_4_1982}. A more general definition of the parameter $\xi_l$ (indicated below as $\xi$) is related to the importance of relativistic effects in the electron transverse motion, with respect to the direction of its average velocity: $\xi=p_{\bot,\max}/mc$, where $p_{\bot}=\gamma mv_{\bot}$ is the transverse momentum \cite{Baier_b_1998}. In the ultrarelativistic regime the parameter $\xi$ also represents the maximum angular deflection during the electron's motion from its average direction, divided by the characteristic angle of radiation $1/\gamma$. In the terminology used in \cite{Baier_b_1998} the conditions $\xi\ll 1$ and $\xi\gg 1$ thus correspond to the dipole and the magnetic bremsstrahlung (or CCF) limits, respectively. 

One of the main reasons why such an old, fundamental and outstanding problem
as the radiation reaction problem is still unsolved, relies on the difficulties 
in detecting it experimentally. As we have hinted above, the rapid development of laser technology 
has renewed the interest in this problem because the strong fields provided by intense laser facilities may
allow for the experimental measurement of radiation-reaction
effects (we refer to the review \cite{Di_Piazza_2012} for papers
until 2012 and we also mention the recent studies \cite{Kravets_2013,
Kumar_2013,PhysRevLett.111.054802,Bashinov_2013b,Ilderton_2013,PhysRevLett.112.015001,Ji_2014,Li_2014,
Capdessus_2015,Vranic_2016,Dinu_2016}). In \cite{Di_Piazza_2017} we 
have realized that the strong electric fields in aligned crystals may be
also suitable for measuring radiation-reaction effects and test
the LL equation. In an aligned crystal, in fact, under suitable 
conditions identifying the so-called channeling regime, an electric 
charge also oscillates similarly as in a laser field and may thus 
radiate a substantial fraction of its energy.

\begin{figure}
\includegraphics[width=1.0\textwidth]{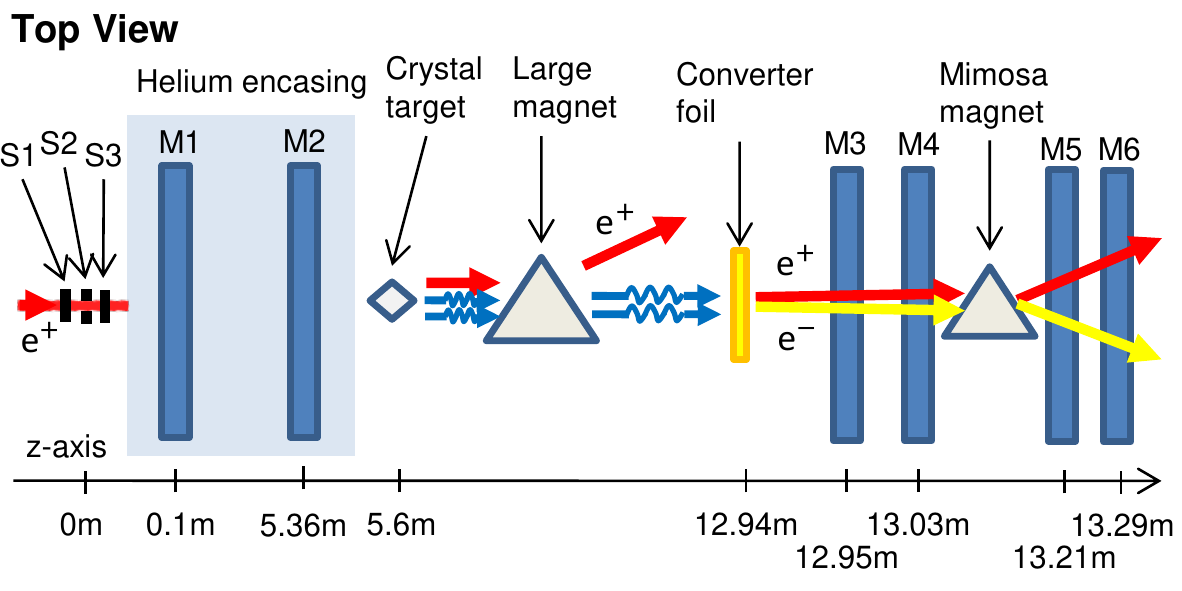}

\caption{Experimental setup. A schematic representation of the experimental setup in the H4 beam line in the SPS North Area at CERN. The symbols ``Sj'', with $j=1,2,3$ denote the scintillators and the symbols ``Mi'', with $i=1,\ldots,6$, denote MIMOSA position sensitive detectors.\label{fig:setupfig}}

\end{figure}

In the experiment described below, ultrarelativistic positrons cross 
a Si crystal in the channeling regime. The dynamics of the positrons is characterized 
by $\chi\leq 1.4$ and $0.7\lesssim\xi\lesssim7$ such that one is in 
the quantum regime and the field is either classically strong or 
in an intermediate regime below the CCF regime $\xi\gg 1$. 
The experiment has been performed at the SPS NA facility at CERN employing
positrons with incoming energy of $\varepsilon_0=178.2\;\mbox{GeV}$ and two
Si crystals with thickness $3.8\;\mbox{mm}$ and $10\;\mbox{mm}$, respectively, 
aligned along the $\langle 111\rangle$ axis. The measured photon emission 
spectra show features which can only be explained theoretically by including 
both 1) quantum effects related to the recoil undergone by the positrons 
in the emission of photons and the stochasticity of photon emission, and 
2) radiation-reaction effects stemming from the emission of multiple photons. 
Several experiments have studied the emission of radiation in crystals in 
the quantum regime, mostly in thin crystals to avoid pile-up effects in 
the calorimeter i.e. the emission of multiple photons by a single particle 
has been avoided. Due to pile-up, in fact, only the sum of the energies 
of all the photons emitted by each charged particle is measured in such experiments, which 
prevents the possibility of reconstructing the single-photon spectrum 
(in e.g. \cite{PhysRevLett.63.2827} such a pile-up effect can be seen). 
In the present experiment we have instead employed a thin converter 
foil and a magnetic spectrometer to obtain the single-photon spectrum, 
see figure \ref{fig:setupfig}. Therefore, in the radiation-reaction 
regime where many photons are emitted by a single positron, the 
current experiment clearly provides more information on the dynamics 
of the positrons and a stronger test of the theory than previous
experimental campaigns.

In figure \ref{fig:setupfig} a schematic of the experimental setup is shown.
The incoming positron encounters the scintillators S1, S2 and S3 which
are used to make the trigger signal. The positron rate is sufficiently low such that in each event only a 
single positron enters the setup. The positron then enters a He
chamber where the two first position sensitive (2 cm $\times$ 1 cm) MIMOSA-26
detectors are placed. Shortly after the He chamber the crystal
target is placed. The He chamber reduces multiple scattering of the positron such
that the incoming particle angle can be measured precisely using the detectors 
M1 and M2. After the positron enters the crystal, multiple photons
and charged particles will leave the crystal. We have ensured also 
theoretically that electron-positron pair production by the emitted 
photons is negligible in the considered experimental conditions. To 
sweep away the charged particles, two large magnets were placed before 
the final set of tracking detectors. The photons emitted inside the 
crystal then reach a thin converter foil, $200\;\text{\textmu m}$ of Ta, 
corresponding to approximately 5\% of the radiation length $X_{0}$
, which in turn corresponds to $7/9$ of the mean free path 
for pair production by a high-energy photon \cite{PhysRevD.86.010001}. The 
thickness was optimized such that most of the time a single photon 
among those emitted by each positron converts to an electron-positron 
pair. The produced pair then passes through M3 and M4 before entering 
a small magnet, such that the momenta of the electron and the positron 
can be determined based on the resulting angular deflection. Finally, 
the deflected electron and positron pass through M5 and M6. As we have mentioned, unlike 
using a calorimeter, this setup has the great advantage that it 
allows one to measure the single-photon radiation spectrum since 
only a single, randomly chosen, of the several emitted photons 
converts to a pair in the thin foil. It is important to point 
out that for photon energies much larger than the electron rest energy, 
as most of those emitted in our experiment, the conversion of a 
photon into an electron-positron pair in the thin foil is independent 
of the photon energy \cite{PhysRevD.86.010001}. Thus, the presence of 
the thin foil does not alter the spectrum of the photons emitted in 
the crystal. The tracking algorithm used in the analysis of the data 
to correctly determine the energy of the photon which originated from 
the measured electron and positron is described in the Supplementary Materials. 
It is clear that the spectrum originating from this procedure can
not be directly compared to the theory since the response of the setup
is complicated by ``practical'' effects such as multiple scattering in 
the converter foil and the presence of air. Therefore a simulation of 
the experimental setup which can ``translate'' the theoretical photon spectra
into the corresponding experimental ones has been developed, the 
details of which can also be found in the Supplementary Materials.

In figure \ref{fig:fig2}, left panel, we show the experimentally obtained counting spectra for 
the ``background'' case, when no crystal is present, for the ``random'' case when 
the crystal is present but not aligned with respect to the positron beam, 
and for the ``align'' case, when the crystal's $\langle 111\rangle$ axis is 
aligned with the positron beam. In the right panel we show a comparison of the experimental and the
theoretical results in the amorphous case. The theoretical, simulated curves, are denoted by ``sim'' (see also the Supplementary Materials). In the vertical label of this 
plot $X_{0}=9.37$ cm is the radiation length of Si. In the random orientation 
the radiation emission is the well understood Bethe-Heitler bremsstrahlung \cite{PhysRevD.86.010001} 
and the agreement here therefore shows that the simulation of the setup is accurate. 
The result in the random 
orientation was used as a way to normalize the theoretical results to the experiment by a scaling factor. 
This is necessary since the efficiency of the setup depends not only on the geometry of
the setup, multiple scattering etc., but also on the inherent efficiency of the MIMOSA detectors.

We have considered four different theoretical models to compare with the experiment. 
These models are described in the section Materials and Methods in the Supplementary 
Materials, and, depending on which effects they include, are indicated as classical 
plus radiation reaction model (CRRM), semiclassical plus radiation reaction model 
(SCRRM), quantum plus radiation reaction model (QRRM), and quantum with no radiation reaction model (QnoRRM). 
In figure \ref{fig:fig3} we show the result of such a comparison in 
the cases of the 3.8mm crystal (left) and the 10.0mm crystal (right).
As we have anticipated, among the four models described in the
Supplementary materials, only the QRRM can be considered in 
reasonable agreement with the experimental data, indicating
the importance of including both quantum and radiation-reaction
effects in the modeling. As we have also hinted, the remaining 
discrepancy can likely be attributed 
to the use of the CCF approximation in regions 
at the limits of its applicability. However, to the best of 
our knowledge, no complete theory of quantum radiation reaction, 
valid in all regimes, has yet been devised as it would essentially 
imply an exact computation of the emission probability of an 
arbitrary number of hard photons.

For the sake of completeness, in figure \ref{fig:fig4} we show the positron power spectra according to the 
four mentioned theoretical models before the translation based on the simulation 
of the setup has been carried out. Here it is seen that for both thicknesses the curves 
corresponding to the ``QnoRRM'' are the same but that this is not the case after 
the translation is carried out (see figure \ref{fig:fig3}). The main reason for this is that the efficiency of the experimental setup depends on the total number of produced photons. 
This effect becomes severe when the number of photons that can convert in the foil 
becomes appreciable compared to $\sim 26$, considering the 5\% of the radiation length $X_{0}$ 
converter foil such that multi-photon conversion becomes likely. In such events the original photon energy 
can not be found and is thus rejected (this also shows the necessity of doing such a simulation of the experimental setup). It is seen in the 3.8 mm case that there is a qualitative agreement between figure \ref{fig:fig3} and figure \ref{fig:fig4} in the relative 
sizes of the spectra compared to each other. However, in the case of the 10.0 mm crystal 
it is seen, for example, that the spectrum corresponding to the ``QRRM'' model is higher than that corresponding to the ``SCRRM'' in figure \ref{fig:fig3}, whereas the opposite occurs in figure \ref{fig:fig4}. This is possible due to the many more soft photons being predicted in the ``SCRRM'' calculation than in the ``QRRM'', which lowers the translated spectrum because of the discussed rejection of multi photon conversion events in the foil.

\begin{figure}
\includegraphics[width=1.0\textwidth]{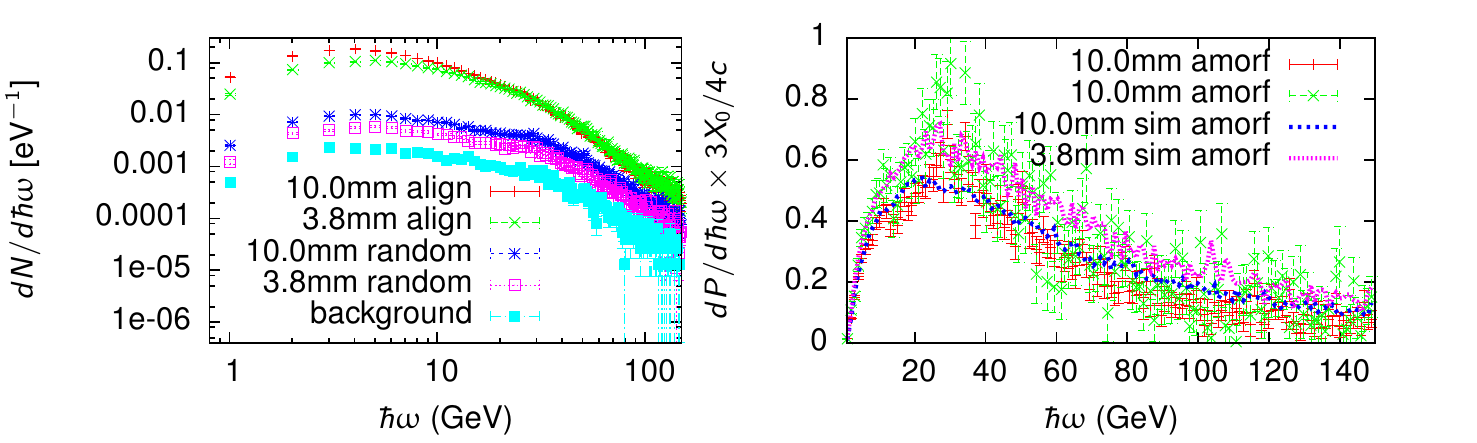}
\caption{Experimental counting spectra and comparison to simulation in random orientation. Photon counting spectra per single incoming positron for the two crystal thicknesses indicated in the text in the aligned and the random case along with a measurement of the background radiation (left panel). The background subtracted power spectrum in the random orientation are compared to simulations (right panel).\label{fig:fig2}}
\end{figure}

\begin{figure}
\includegraphics[width=1.0\textwidth]{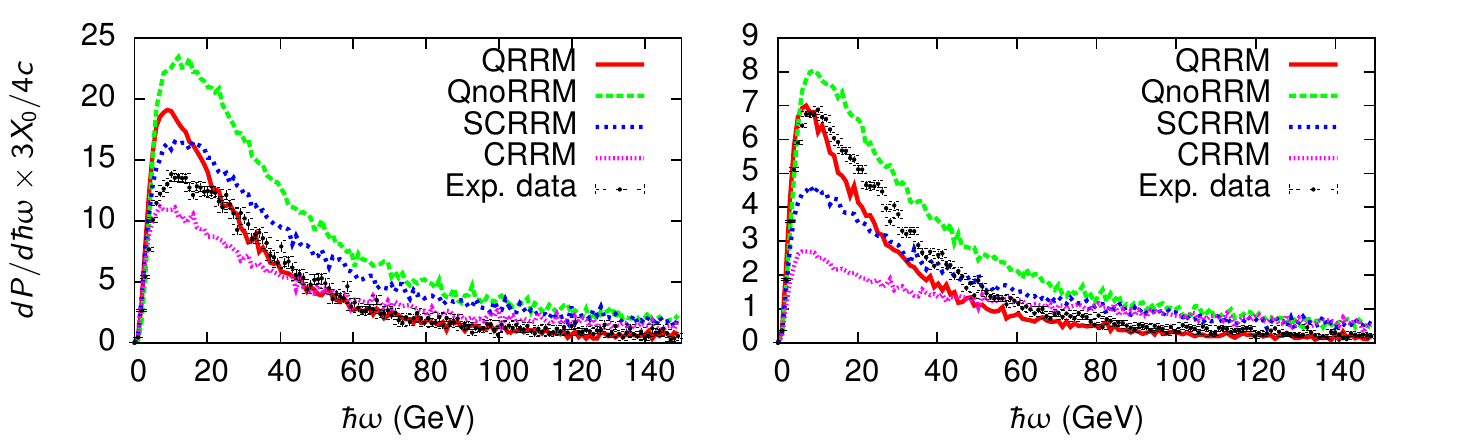}
\caption{Experimental power spectra. Background subtracted power spectra in the aligned case for two crystal thicknesses: 3.8 mm (left panel) and 10.0 mm (right panel). The experimental data are compared to the four different theoretical models described in the Supplementary Materials after being translated through the simulation of the experimental setup.\label{fig:fig3}}
\end{figure}

\begin{figure}
\includegraphics[width=1.0\textwidth]{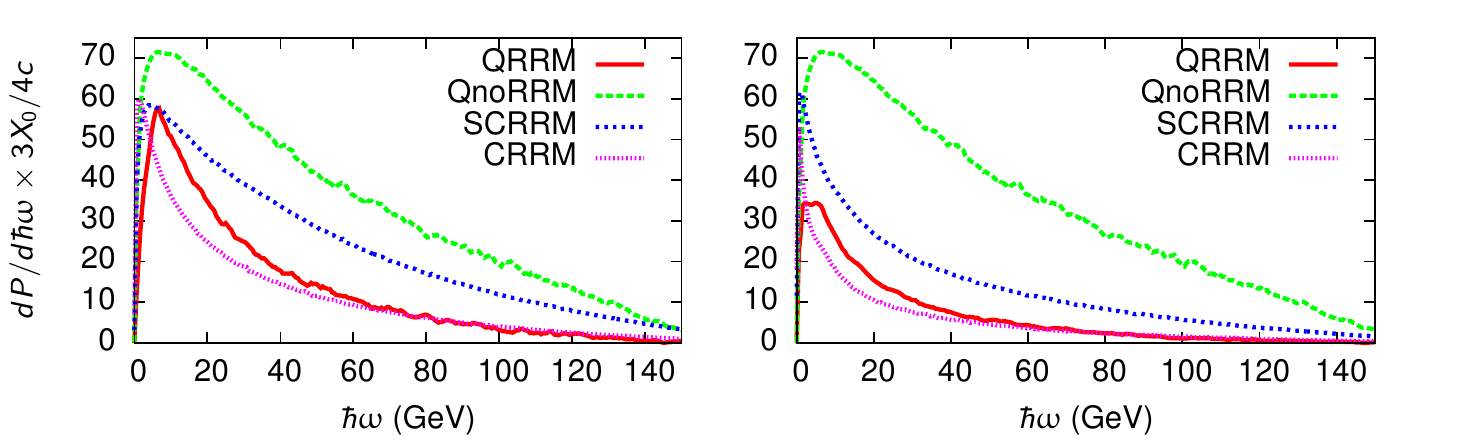}
\caption{Theoretical power spectra. Power spectra calculated according to the four different theoretical models described in the Supplementary Materials for the two crystal thicknesses: 3.8 mm (left panel) and 10.0 mm (right panel).\label{fig:fig4}}
\end{figure}


\section*{Supplementary Materials and Methods}

\textbf{Theoretical Models} We have considered four different theoretical models to compare with the experiment:

\begin{enumerate}
\item Classical plus radiation reaction model (CRRM). In this model, we include radiation reaction classically, i.e., we determine the positron trajectory via the Landau-Lifshitz (LL) equation $mcdu^i/ds=(q/c)F^{ij}u_j+g^i$, where (see, e.g., \cite{Landau_b_2_1975})
\begin{equation}
g^{i}=\frac{2q^{3}}{3mc^3}\frac{\partial F^{ik}}{\partial x^{l}}u_{k}u^{l}-\frac{2q^{4}}{3m^{2}c^5}F^{il}F_{kl}u^{k}+\frac{2q^{4}}{3m^{2}c^5}(F_{kl}u^{l})(F^{km}u_{m})u^{i}.\label{eq:Llforce-1}
\end{equation}
Here, $u^i$ is the positron four-velocity, $s$ its proper-time, $q=e>0$ and $m$ its charge and mass respectively, and $F^{ij}$ the electromagnetic field tensor of the crystal (see, e.g., \cite{Jackson_b_1975}). The crystal field has been modeled starting from from the sum of Doyle-Turner string potentials \cite{Moller1995403}, centered on a regular grid according to the diamond cubic crystal structure. Once the positron trajectory has been determined, the emission spectrum has been computed starting from the Li\'{e}nard-Wiechert potential and following the standard procedure, as described, e.g., in \cite{Jackson_b_1975}.
\item Semiclassical plus radiation reaction model (SCRRM). In this model we ``partially'' include quantum effects following an approach described, e.g., in \cite{PhysRevE.81.036412}, where the term involving the derivative in Eq. (\ref{eq:Llforce-1}) is neglected and the remaining two are multiplied by the ratio between the quantum total emitted power and the corresponding classical quantity. The emission spectrum has then been evaluated as in the CRRM. This model phenomenologically takes into account that quantum effects reduce the total radiation yield but it does not account for the intrinsic stochasticity of the photon emission process (see, e.g., \cite{Neitz_2013}).

\item Quantum plus radiation reaction model (QRRM). In this model the radiation emission is taken into account fully quantum mechanically, i.e., the positron propagates classically within the crystal according to the Lorentz equation and, in a genuinely random way, it emits photons and undergoes the corresponding recoil. A numerical program has been written to determine the positron dynamics and emission spectrum according to the following procedure. At each step in the time evolution of the positron trajectory the probability of single photon emission in that time interval is originally calculated within the constant-crossed field (CCF) approximation (see, e.g., Eq. (4.36) in \cite{Baier_b_1998}) and a random number generator decides if the emission takes place or not (the time step has been chosen sufficiently small such that the resulting single-photon emission probability is much smaller than unity). In the former case, on the one hand, the photon energy is also determined by sampling another random number using the procedure shown in \cite{Yokoya1985ab}, such that the emitted photons are consistently  distributed in accordance with the formula for radiation emission also in the CCF approximation (see Eq. (4.24) in \cite{Baier_b_1998}). On the other hand, the momentum of the emitted photon is directed along the positron momentum at the instant of emission (which is an excellent approximation in the ultrarelativistic regime) and it is subtracted from the positron momentum before the trajectory solver starts out with the new initial conditions according to the Lorentz equation. As we have mentioned, the emission probabilities within the CCF approximation are employed here (like in e.g. \cite{Neitz_2013,PhysRevLett.112.015001,PhysRevLett.105.220403}). However, since $\xi$ is in some cases only comparable to unity in the experiment, the model has been improved. In fact, for low energies of the emitted photons $\hbar\omega\ll \varepsilon_0$, the formation length of the emission process is given by $l_{f}(\omega)=2\gamma^{2}c/\omega$, where $\gamma$ is the Lorentz factor of the positron at the instant of emission \cite{Baier_b_1998}. Thus, if we denote as $\lambda_{0}$ the typical oscillation length of a channeled positron, we expect that the CCF approximation does not work for frequencies lower than $\omega_c$, with $l_{f}(\omega_c)=a\lambda_{0}/2$, where $a$ is a constant of the order of unity. In order to determine the constant $a$, we have used the fact that at low photon energies, where it is not valid, the CCF approximation significantly overestimates the emitted radiation yield with respect to the more general and more accurate approach of Baier et. al. described in \cite{Baier_b_1998,Belkacem198586,PhysRevD.90.125008,PhysRevD.92.045045} and also used numerically in several channeling codes \cite{Bandiera201544,PhysRevA.86.042903}. Thus, as the simplest approach, we have modified the CCF emission probability by setting it to zero for photon frequencies below $\omega_c$. The constant $a$ has been fixed by requiring that the resulting total yield coincides with the total yield given by the more accurate approach by Baier et al. \cite{Baier_b_1998,Belkacem198586,PhysRevD.90.125008,PhysRevD.92.045045} in the case of a thin crystal where multiple photon emissions are negligible. Indeed, we have found numerically that in this way $a$ turns out to be approximately equal to 0.52.
\item Quantum with no radiation reaction model (QnoRRM). This model is the same as the QRRM described above but whenever the positron emits a photon, its recoil energy-momentum is not subtracted from the positron. The spectrum in the QRRM approaches the spectrum of this model when the crystal becomes thin because for a thin crystal the probability of multiple photon emission becomes negligible and each positron essentially emits a single photon. Thus, the difference between this model and the QRRM shows the size of the effects of the recoil of multiple photon emission, i.e. of radiation reaction.
\end{enumerate}

\noindent \textbf{Tracking Algorithm} A tracking algorithm has been employed 
in the analysis of the experimental data in order to correctly characterize 
the created electrons and positrons, and determine 
whether they arise from a converting photon in the foil. 
This is decided based on a series of conditions: Hypothetical rectilinear tracks 
in the detectors M3-M4 and M5-M6 (see figure 1 in the main text) are 
constructed by connecting all possible pairs of hits in the two 
planes of M3-M4 and in the two planes of M5-M6. These track 
candidates in M5-M6 must be matched with those 
in M3-M4 giving a full particle track, identified by the following conditions:
\begin{enumerate}
\item The tracks for individual particles arising from two points
in M3-M4 and from two points in M5-M6 are ideally
continued into the magnet and, in order to be accepted, they must have a 
distance to each other within 0.8 mm in the center of the magnet. 
\item The two tracks from the detectors M3-M4 and M5-M6 should be at the
shortest distance from each other approximately at the $z$ position
of the center of the magnet.
\item The size of the deflection angle between the tracks in M5-M6
and the tracks in M3-M4 in the $y$ direction must be smaller
than 2 mrad because the magnet deflects only along the $x$ direction. 

\end{enumerate}

Now, tracks of electrons and positrons have been individually 
identified. Moreover, these must also be paired to stem from 
the same photon. This identification is carried out by 
requiring that an electron and a positron track 
must originate from within a distance of 20 \textmu{}m on
the $x$-$y$ plane in the converter foil. 
After the identification of the tracks has been carried out, it may happen that 
for a given electron or positron, more than one particle of 
opposite charge matches within the mentioned distance in the
converter foil. If this happens, the event is discarded because 
more than one photon must have converted in the foil and it is not
possible to unequivocally associate the electron-positron pair with a photon.

This also implies that if the number of photons above the pair production 
threshold in the converter foil exceeds $\sim 26$, one will begin to see 
the experimental photon spectrum drop due to multiple photon 
conversion.
We recall that, as we have mentioned in the main text, 
the thickness of the converter foil corresponds to about $(7/9)\times 
5\%\simeq 1/26$ of the average length that a photon covers before 
converting into an electron positron pair. Therefore, optimally, this regime is avoided. 

In each event all tracks are determined in 
M1, M2, and M3 as well. The chosen track in these detectors is the 
one with the closest approach to the pair origin already 
determined in the converter foil. Finally, the positron entry angle 
is determined from the hits in M1 and M2 of this track.

It is clear that the photon energy spectrum originating from this procedure can
not be directly compared to the theoretical spectra because the response of this 
setup is complicated by practical issues. For example, a positron entering the 
setup at the center of the detector M1 and another one entering at the border 
of the same detector will have a different chance of leading to a detected pair. 
The reason is that the pair originating from the positron hitting M1 at the border 
is more likely to be deflected outside M5 or M6. A similar effect takes place 
when considering the angle of the incoming positrons. In addition 
to this, multiple scattering in the converter foil, air and detectors 
influence both the efficiency and the resolution. In order to deal with 
these issues, a code simulating the setup has been written. The beam distribution 
in position and angle as experimentally measured are given as input to the program 
simulating the setup, and then the effects of multiple Coulomb scattering between 
and inside the detectors and converter foil, as well as of the Bethe-Heitler
pair-production are included for determining the particles' dynamics 
\cite{PhysRevD.86.010001}. The only non-trivial input to such a simulation 
of the setup, is the spectrum of the radiation emitted by the positrons
in the crystal, which we have determined theoretically according to the four 
models described above. Finally, the simulation of the setup produces data-files 
of the same format as those obtained from the data acquisition in the experiment, 
which are then both sent through the tracking algorithm.

\section*{Acknowledgments}
We acknowledge the technical help and expertise from Per Bluhme Christensen,
Erik Loft Larsen and Frank Daugaard (AU) in setting up the experiment
and data acquisition.

\section*{Authors' contribution}
TNW and UIU conceived and carried out the experiment. TNW and ADP carried out the theoretical calculations. TNW carried out the data analysis. TNW and ADP wrote the paper with input and discussion from HVK and UIU.

\end{document}